\documentclass[review]{elsarticle}

\usepackage[utf8]{inputenc}
\usepackage{hyperref}
\usepackage{amsmath, amsfonts}
\usepackage{bm}    
\usepackage{physics}
\usepackage{siunitx}
\journal{Optik}

\begin{document}

\begin{frontmatter}

\title{Phonon-assisted tunnelling in a double quantum dot molecule immersed in a cavity}

%% Group authors per affiliation:
\author[un]{Vladimir Vargas-Calderón}
\ead{vvargasc@unal.edu.co}
\author[un]{Herbert Vinck-Posada}
\ead{hvinckp@unal.edu.co}
\address[un]{Departamento de Física, Universidad Nacional de Colombia, AA 055051, Bogotá, Colombia}

\begin{abstract}
The effects caused by phonon-assisted tunnelling (PhAT) in a double quantum dot (QD) molecule immersed in a cavity were studied under the quantum Markovian master equation formalism in order to account for dissipation phenomena. We explain how for higher PhAT rates, a stronger interaction between a QD and the cavity at off-resonance takes place through the resonant interaction of another QD and the cavity, where the QDs interact through tunnelling. A closer look at the system's allowed optical transitions as a function of PhAT rates showed coalescence of the transition energies, leading to a loss of spectroscopic resolution in the photoluminescence (PL) spectrum. This coalescence occurs at several exceptional points, suggesting a quantum dynamical phase transition.
\end{abstract}


\end{frontmatter}

\section{Introduction}

Quantum dot architectures remain to be a promising candidate for the implementation of quantum computing \cite{steffen2011,kloeffel2013}. Both lateral electric gate-based or self-assembled QDs are attractive physical systems to be considered as elementary blocks of a quantum computer (among many other applications) and also can be built through lithographic techniques that have already been used by the chip building industry for large-scale integration into more complex architectures. Moreover, the populations and transitions of such systems can be controlled by immersing them in a cavity able to confine an electric field. These sort of light-matter interaction experiments have been carried out in semiconductor quantum dots with micropillars~\cite{Reithmaier2004}, photonic crystals~\cite{Yoshie2004}, microdisks~\cite{peter2005}, among other cavities; and in lateral quantum dots with superconductor resonators as cavities ~\cite{kulkarni2014,stockklauser2017}. 

The study of elementary devices for quantum computing is necessary to characterise their reaches and limitations. In the case of double quantum dot (DQD) molecules, which can be employed as qubits~\cite{Nichol2017,friesen2003,harvey2017}, the quantum tunnelling effect can be used to transfer excitations (excitons or electrons) from a quantum dot to the other. Clearly, uncontrolled tunnelling must be avoided to keep the information as is. The phonon-assisted tunnelling effect is regarded as an important source of tunnelling, so that a thorough study is needed to understand its effects. So far, it has been observed both in semiconductor QDs~\cite{berghoff2008} and in lateral QD molecules~\cite{braakman2013}. The origin of PhAT is the Coulomb interaction between an excitation of a QD molecule and the surrounding lattice, and occurs when the excitation transfer implies an energy mismatch, which is compensated by the emission or absorption of a phonon, so that an important portion of PhAT described by the phonon spectral density~\cite{grodecka2008}.

In this work we study the effect of PhAT onto the main observables of a DQD immersed in a unimodal electric field confined by a cavity. To do so, in section \ref{sec:QME} we introduce the quantum master equation formalism that effectively describes the dynamics of the open quantum system at hand; in section \ref{sec:emission} we describe the way of computing the PL spectrum of such a system, and we set the Liouvillian spectral decomposition as a tool for understanding the spectrum, and all the transitions in the system; in section \ref{sec:results} we discuss the main results of our research; and finally we present some conclusions in section \ref{sec:conc}.

\section{Modelling of DQD-cavity system\label{sec:QME}}

We model the DQD-unimodal cavity system as two two-level systems and a quantum harmonic oscillator, respectively, with a Hamiltonian
\begin{align}
H = \sum_{j=1}^2\hbar\omega_j\sigma_j^\dagger\sigma_j + \hbar\omega_0a^\dagger a + \hbar T (\sigma_1^\dagger\sigma_2 + \sigma_1\sigma_2^\dagger) + \sum_{j=1}^2 g_j(\sigma_j^\dagger a + \sigma_j a^\dagger)
\end{align}
where the first two terms correspond to the energy of the QDs and the cavity mode, the third term is an effective tunnelling between the QD1 and QD2~\cite{jefferson1996} and the fourth term contains the Jaynes-Cummings interaction between each QD and the cavity mode~\cite{jaynes-cummings1963}. Here $\sigma_j^\dagger (\sigma_j)$ is the creation (annihilation) operator for the $j$-th QD with frequency $\omega_j$, and $a^\dagger (a)$ is the creation (annihilation) operator for the cavity mode with frequency $\omega_0$, $T$ is the tunnelling rate and $g_j$ is the cavity quantum electrodynamical coupling constant between the mode and the $j$-th QD.

The DQD-cavity system is an open quantum system subjected to dissipative processes such as incoherent pumping and spontaneous emission for each QD and also escape and pumping of photons for the cavity~\cite{tejedor2004}. Pure dephasing of each QD can be safely neglected at low temperatures. In order to isolate the PhAT effect, we do not consider the phonon assistance that compensates for the energy mismatch in the excitation-photon transitions~\cite{majumdar2011,hohenester2010}. Finally, we introduce PhAT by proposing a linear interaction between the DQD tunnelling term $\sigma_1^\dagger\sigma_2 + \sigma_1\sigma_2^\dagger$ and a phonon bath $\sum_k \hbar\omega_k b^\dagger_k b_k$ similar to~\cite{bagheri2012,rozbicki2008,karwat2011,Karwat2014}
\begin{align}
H_I = (\sigma_1^\dagger\sigma_2 + \sigma_1\sigma_2^\dagger)\sum_k (g_k b_k + g_k^* b_k^\dagger).
\end{align} 
Therefore, in the case of the Born-Markov approximation and approximating quasi-degenerate eigenenergies of the Hamiltonian as degenerate ones, the system evolves under an effective quantum master equation of the form~\cite{breuer2007}
\begin{align}
\dot{\rho}(t) = -\frac{i}{\hbar}[H, \rho(t)] + \sum_{j=1}^2(\gamma_j \mathcal{D}_{\sigma_j} + P_j\mathcal{D}_{\sigma_j^\dagger})   +  P\mathcal{D}_{a^\dagger} + \kappa \mathcal{D}_{a} + P_T \mathcal{D}_{\sigma_1^\dagger\sigma_2 } + \gamma_T \mathcal{D}_{\sigma_1\sigma_2^\dagger} ,\label{eq:qme}
\end{align}
where $P(\gamma)_j$ is the pumping(spontaneous emission) rate on the $j$-th QD, $P(\kappa)$ is the pumping(escape) rate of photons to(from) the cavity and $P(\gamma)_T$ is the PhAT pumping(decay) rate. The dissipation superoperators in Lindblad form are defined by $\mathcal{D}_O =  O\rho(t)O^\dagger - \frac{1}{2}\{O^\dagger O, \rho(t)\}$~\cite{breuer2007}.

\section{Photoluminescence spectrum and Liouvillian spectral decomposition\label{sec:emission}}
An important measurement in these types of systems is the PL spectrum, because it allows to probe strong coupling through the presence of anti-crossing regions and Rabi splitting. Through the Wiener-Khintchine theorem it is possible to obtain a formula to compute the PL spectrum~\cite{tejedor2004,scully_zubairy_1997,mollow1969,edgar2016} which reads
\begin{align}
I(\omega) \propto \frac{\kappa}{\pi}\lim_{t\to\infty}\int_{0}^\infty \expval{a^\dagger(t)a(t+\tau)}e^{i\omega\tau} d\tau,\label{eq:spectrum}
\end{align}
where the two-point correlation function in the integrand of Eq.~\eqref{eq:spectrum} can be obtained through the Quantum Regression Theorem by first computing the single-operator expected values~\cite{breuer2007}.

In order to understand the transitions that produce the PL spectrum, a closer look at the Liouvillian is indispensable. At any time $t$, the density matrix of the system is $\rho(t) = \exp(\mathcal{L}t)\rho(0)$, where $\mathcal{L}$ is the Liouvillian superoperator defined by $\dot{\rho}(t) = \mathcal{L}\rho(t)$, as in Eq.~\eqref{eq:qme}. In other words, the density matrix of the DQD-cavity system satisfies 
\begin{align}
\rho(t) = \sum_k c_k \exp(\Lambda_kt) \varrho_k,\label{eq:evolution_rho}
\end{align} 
where $c_k$ are constants that satisfy the initial condition and $\{\varrho_k,\Lambda_k\}$ conform the right eigensystem of $\mathcal{L}$, i.e. $\mathcal{L}\varrho_k = \Lambda_k\varrho_k$. Therefore, the system will only occupy states $\rho$ spanned by the eigenmatrices $\{\varrho_k\}$.

Since $\mathcal{L}$ is not necessarily Hermitian, the eigenvalues $\{\Lambda_k\}$ are complex in general. Such eigenvalues have the information of both the frequencies (in the imaginary part) as well as the linewidths (in the real part) of the allowed transitions in the system~\cite{albert2014}. This can be checked by taking the real part of the Fourier transform of each time signal in Eq.~\eqref{eq:evolution_rho}, i.e. $\Theta(t)\exp([\mathfrak{Re}(\Lambda_k) + i\mathfrak{Im}(\Lambda_k)]t)$, where $\Theta$ is the Heaviside step function.

In order to isolate the transition frequencies and linewidths that are optical and therefore related to the PL spectrum, one must realise that optical transitions happen in transitions between states belonging to consecutive varieties of excitation. It has been shown in Ref.~\cite{torres2014} that the gain and loss dissipative processes mix all the varieties of excitation, but when ignoring gain or loss dissipative processes, the Liouvillian can be written in block diagonal form, where each block couples pairs of varieties of excitation in a closed fashion. Therefore, the frequency of optical transitions can be obtained by solving the eigensystem~\cite{santiEcheverri2018}
\begin{align}
\mathcal{L}^{n,n-1}\varrho_k^{n,n-1}=\lambda_k^{n,n-1}\varrho_k^{n,n-1},
\end{align}
where $\mathcal{L}^{n,n-1}$ is the subspace, or block, that contains the information of the two consecutive varieties of excitation $n$ and $n-1$.

\section{Results\label{sec:results}}
We consider realistic parameters for the DQD-cavity system by examining the work in Ref.~\cite{laucht2010}, where the exciton energies in the QDs are approximately $\hbar\omega_1=\hbar\omega_2=1.218\si{\electronvolt}$ and the cavity mode has the same frequency $\omega_0$. Additionally, they report QD-cavity coupling constants $\hbar g_1=44\si{\micro\electronvolt},\hbar g_2 = 51\si{\micro\electronvolt}$, excitonic pumping to the QDs $\hbar P_1=1.5\si{\micro\electronvolt}, \hbar P_2=1.9\si{\micro\electronvolt}$, spontaneous emission of $\hbar\gamma_1=0.1\si{\micro\electronvolt}, \hbar\gamma_2 = 0.8\si{\micro\electronvolt}$, escape of photons of $\hbar\kappa=147\si{\micro\electronvolt}$, and pumping to the cavity of $\hbar P = 5.7\si{\micro\electronvolt}$. We consider the weak tunnelling regime~\cite{villasboas2010} by setting $\hbar T = 10\si{\micro\electronvolt}$.

Although Ref.~\cite{laucht2010} mentions that they achieved strong coupling, we consider from now on $\hbar g_1=440\si{\micro\electronvolt},\hbar g_2 = 510\si{\micro\electronvolt}$, so that the Jaynes-Cummings interaction is clearly dominant over the escape of photons dissipative term, which guarantees strong coupling in our calculations.

\begin{figure*}[!ht]
\includegraphics[width=\textwidth]{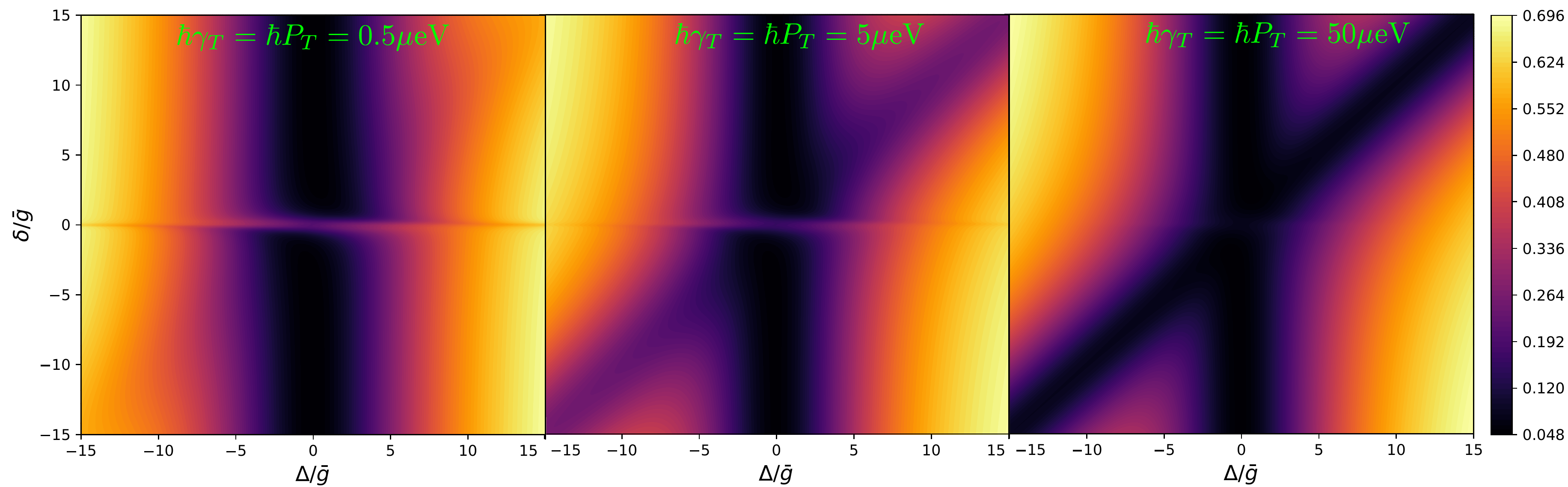}
\caption{Mean excitations of QD1 as a function of interdot detuning $\delta$ and QD1-cavity detuning $\Delta$ for different values of PhAT rates. Axes are normalised to the scale of the Jaynes-Cummings interaction constant $\bar{g} = (g_1+g_2)/2$.\label{fig:3plots}}
\end{figure*}

With this in mind, we demonstrate in Fig.~\ref{fig:3plots} that one effect of PhAT on a DQD-cavity system is that a strong interaction between a QD and the cavity might be achieved even when this QD and the cavity are not in resonance because the other QD mediates such interaction. In the case of Fig.~\ref{fig:3plots}, the mean excitation of QD1 is shown in the stationary case for different cavity-QD1 detuning values $\delta = \omega_0-\omega_1$ and QD1-QD2 detuning values $\Delta=\omega_2-\omega_1$. The line $\Delta=0$ defines the resonance between QD1 and the cavity mode, $\delta=\Delta$ defines the resonance between QD2 and the cavity mode, and $\delta = 0$ is the resonance between both QDs.

At low PhAT rates (left panel of Fig.~\ref{fig:3plots}), the mean excitations in QD1 depend strongly on $\Delta$. When $\Delta=0$, Rabi oscillations are expected between QD1 and the cavity. Since the escape of photons from the cavity is the main dissipative channel, excitations transferred to the cavity from QD1 are expected to be lost. This explains the low mean excitations at the QD1-cavity resonance. It must be noticed that there exists a region at $\delta=0$ where a sudden increase in the mean excitations is located. The interdot resonance causes large oscillations of the excitation between both QDs: the smaller the interdot detuning, the larger the population transfer between the QDs. Since $\delta=0$ maximises the excitation transfer between the QDs, we believe that these oscillations protect the excitation in the QDs from the interaction with the cavity by delocalising the excitation, leading to the aforementioned sudden increase in the mean excitation of QD1.

As PhAT rates increase (central and right panels of Fig.~\ref{fig:3plots}), QD1 depletion from excitations also increase at the resonance between QD2 and the cavity, even when $\Delta$ is large. This means that PhAT allows a strong interaction between QD1 and the cavity at the resonance between QD2 and the cavity, far from the resonance between QD1 and the cavity. If there were ways of controlling PhAT magnitude, this can be advantageous to control the population in a QD which is not resonant with the cavity. Furthermore, since PhAT dampens the interdot oscillations of the populations, the line found at $\delta=0$ blurs away as PhAT rates increase.

\begin{figure}[!ht]
\includegraphics[width=\columnwidth]{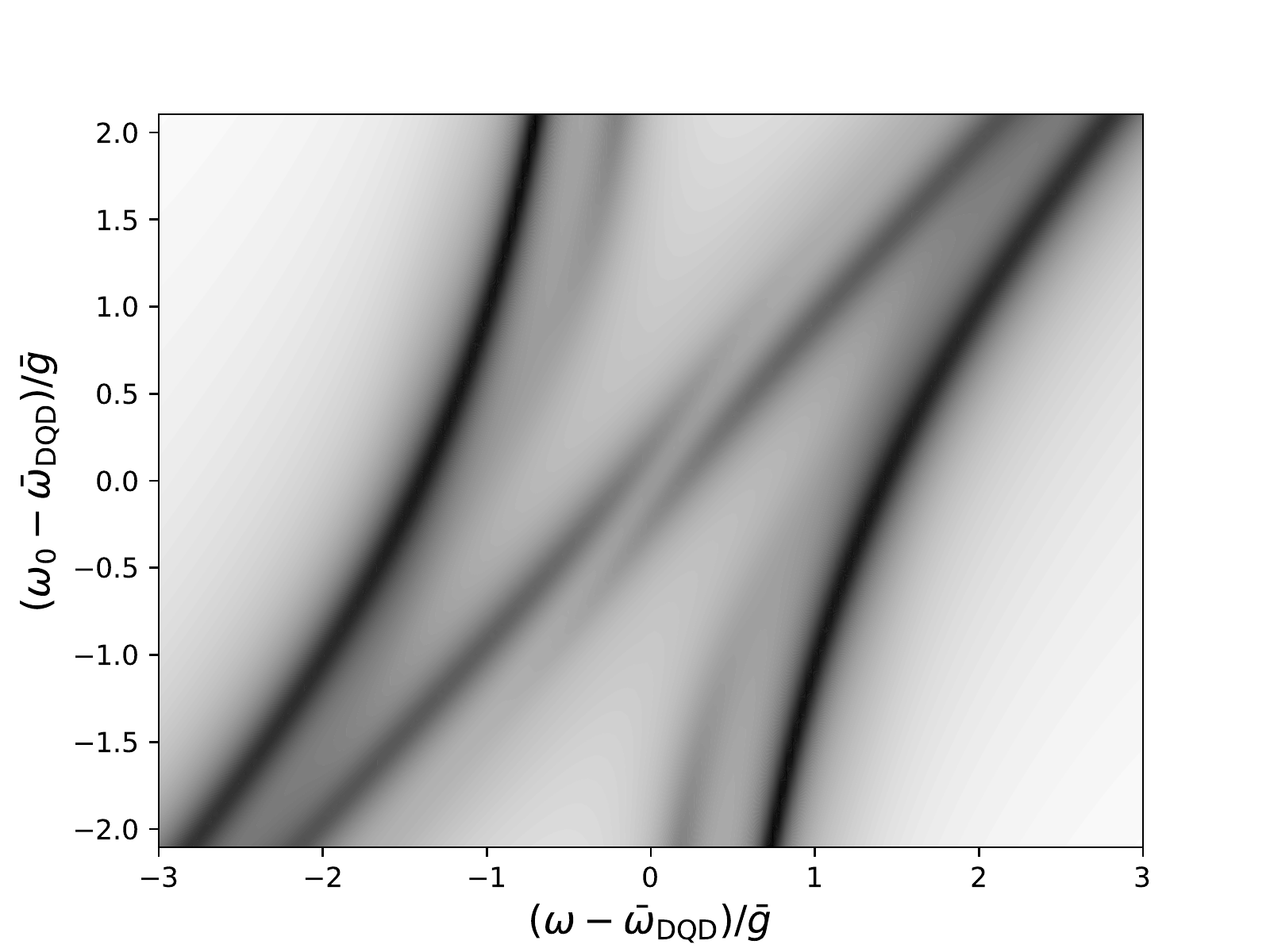}
\caption{PL spectrum as a function of cavity detuning with respect to $\bar{\omega}_\text{DQD}=(\omega_1+\omega_2)/2$ at $\hbar\gamma_T = \hbar P_T = 0.5\si{\micro\electronvolt}$.\label{fig:spec_det}}
\end{figure}

On the other hand, by looking at the PL spectrum as a function of detuning between the cavity and the DQD, we corroborate the strong coupling behaviour, expressed as anti-crossings and Rabi splitting, as shown in Fig.~\ref{fig:spec_det}. The diagonal peak is located at the cavity frequency and presents a splitting due to the difference in the QD-cavity coupling constants. Such a peak does not appear in the presence of only one QD coupled with the cavity. Moreover, there is a coexistence of strong and weak lateral peaks. The strong ones exhibit the anti-crossing behaviour of a QD-cavity system in the strong coupling regime. In this system, when the dissipative rates are very small, weaker split pairs of peaks appear, corresponding to optical transitions of higher varieties, but they depend linearly on the detuning, showing no coupling at all. However, when two QDs interact strongly with light, as is the case of the DQD-cavity system under study, such weak peaks also present an anti-crossing, exhibiting further strong coupling.

\begin{figure}[!ht]
\includegraphics[width=\columnwidth]{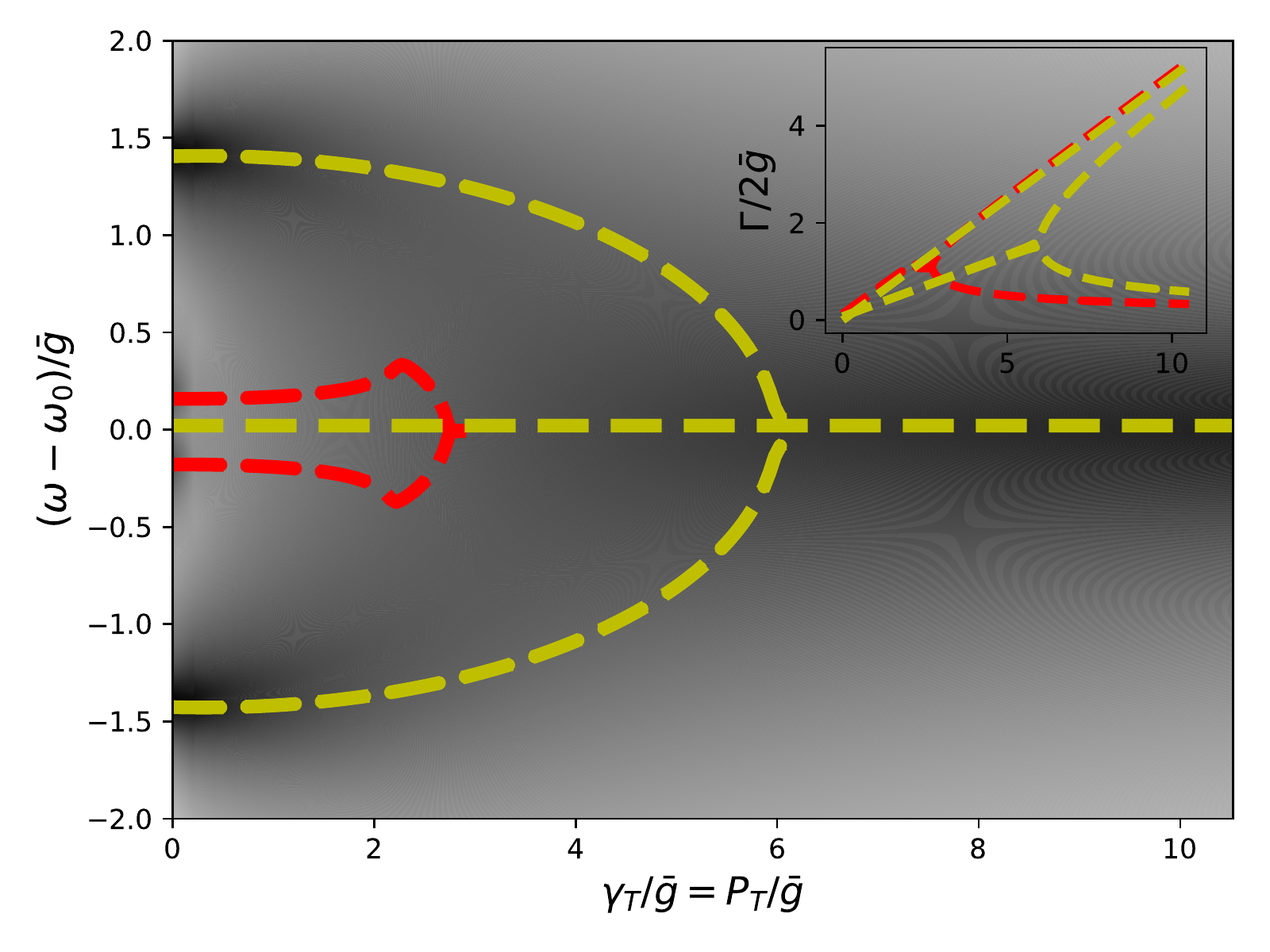}
\caption{PL spectrum as a function of PhAT rates. Lines correspond to transition frequencies between varieties $0\leftrightarrow 1$ (yellow) and $1\leftrightarrow 2$ (red), and the inset shows their corresponding linewidths $\Gamma$.\label{fig:spec_phat}}
\end{figure}

Finally, the PL spectrum as a function of PhAT rates magnitude is shown in Fig.~\ref{fig:spec_phat}. For very low values of PhAT rates, four peaks can be distinguished (c.f. Fig.~\ref{fig:spec_det} at $\omega_0=\bar{\omega}_\text{DQD}$). It looks like the central peaks rapidly broaden and coalesce, while the exterior peaks broaden and merge into one central peak that does not broaden as PhAT rates increase. Since the escape of phonons is the main dissipation channel, the system is expected to always occupy the lowest varieties of excitation. Therefore, by studying the transition frequencies and linewidths via the Liouvillian spectral decomposition, we are able to explain the PL spectrum in detail. The yellow lines show the transition frequencies between varieties 0 and 1, and the red ones between 1 and 2. When the frequencies related to $0\leftrightarrow 1$ transitions coalesce, their linewidths bifurcate: one remains bounded and the other one grows, as depicted in the inset of Fig.~\ref{fig:spec_phat}. The same happens with the exterior lines of frequencies related to $1\leftrightarrow 2$ transitions. Those emission lines with huge linewidths cannot be resolved, whereas the ones with bounded linewidths compose the actual PL spectrum.

\begin{figure}[!ht]
\includegraphics[width=\columnwidth]{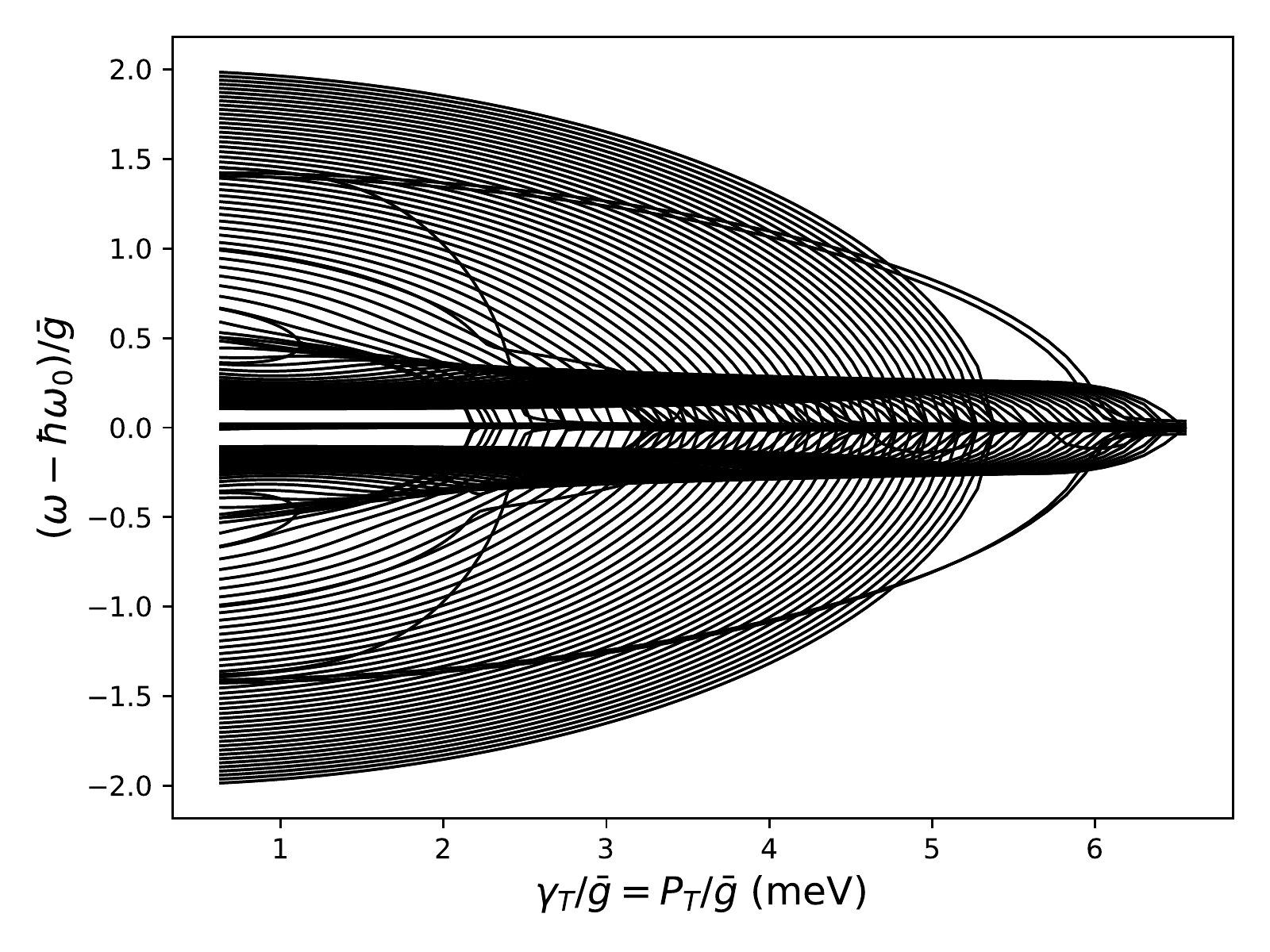}
\caption{Dependence of transition frequencies up to variety of excitation 50 on PhAT rates.\label{fig:transition_freqs}}
\end{figure}

The coalescence behaviour of the PL spectrum and the transition frequencies found for the first varieties of excitation shown in Fig.~\ref{fig:spec_phat} is a general behaviour that occurs in other varieties of excitation which not reachable by the system due to huge photon losses in the cavity. However, the Liouvillian spectral decomposition analysis allows us to uncover those transitions and their behaviour as a function of PhAT as seen in Fig.~\ref{fig:transition_freqs}. A general trend is clear: the system loses spectroscopic individuality, caused by the coalescence of transition frequencies, where pairs of transition frequencies merge at exceptional points~\cite{rotter2010}. This behaviour has been found before in a study of phonon cavity feeding~\cite{santiEcheverri2018}, and hinters a quantum dynamical phase transition~\cite{heyl2018,eleuch2013}.

\section{Conclusions\label{sec:conc}}
We showed how PhAT can greatly increase the interaction between a QD and a unimodal cavity at off-resonance through a second QD at resonance with the cavity. This can be advantageous for applications in the area of high speed logic devices, but can also lead to undesired tunnelling. It remains an important issue to understand how can we experimentally produce and control phonon baths that shape PhAT.

Also, the Liouvillian spectral decomposition allowed us to understand PL spectra through the allowed transitions in the DQD-cavity system, and a general overview of such transitions as a function of PhAT rates exhibited exceptional points, which are clear signatures of a quantum dynamical phase transition.

\section*{Acknowledgements}
The authors acknowledge partial financial support from COLCIENCIAS under the project “Emisión en sistemas de Qubits Superconductores acoplados a la radiación. Código 110171249692, CT 293-2016, HERMES 31361”, and project "Control dinámico de la emisión en sistemas de qubits acoplados con cavidades no-estacionarias, HERMES 41611".

\section*{References}

\bibliography{mybibfile}

\end{document}